\pdfoutput=1

\documentclass[11pt]{article}

\usepackage{authblk}

\usepackage[preprint]{acl}

\usepackage{times}
\usepackage{latexsym}
\usepackage{booktabs}

\usepackage[T1]{fontenc}

\usepackage[utf8]{inputenc}

\usepackage{microtype}

\usepackage{inconsolata}

\usepackage{graphicx}

\usepackage{multirow}

\title{Domain-Specific Query Understanding for Automotive Applications: A Modular and Scalable Approach}

\author[ ]{Isha Motiyani}
\author[ ]{Abhishek Kumar}
\author[ ]{Tilak Kasturi}

\affil[ ]{Predii}

\affil[ ]{\texttt{isha.motiyani@predii.com}, \texttt{abhishek.kumar@predii.com}, \texttt{tilak@predii.com}}

\begin{document}
\maketitle
\begin{abstract}
Despite the growing prevalence of large language models (LLMs) in domain-specific applications, the challenge of query understanding in the automotive sector still remains underexplored. This domain presents unique complexities due to its specialized vocabulary and the diverse range of user intents it encompasses. Unlike general-purpose assistants, automotive systems must precisely interpret user queries and route them to appropriate underlying tool, each designed to fulfill a distinct task such as part recommendations, repair procedures, or regulatory lookups. Moreover, these systems must extract structured inputs precisely aligned with the schema required by each tool. In this study, we present a novel two-step system for domain-specific query interpretation in the automotive context that achieves an effective balance between responsiveness, reliability, and scalability. Our initial single-step approach, which jointly performed classification and entity extraction, exhibited moderate performance and higher latency. By decomposing the task into a lightweight classification stage followed by targeted entity extraction using smaller, specialized prompts, our system achieves substantial gains in both efficiency and accuracy. Due to the niche nature of the automotive domain, we also curated a high-quality dataset by combining manually annotated and synthetically generated samples, all reviewed by domain experts. Overall, our findings demonstrate that decomposing query understanding into modular subtasks leads to a scalable, accurate, and latency-efficient solution. This approach establishes a strong ground for practical deployment in real-world automotive query understanding systems.
\end{abstract}

\section{Introduction}

In real-world query understanding systems—especially in domain-specific settings like automotive support, user queries must be accurately routed to one of many specialized internal tools, each engineered to address a distinct category of information need. This study focuses on the dual challenge of classifying domain-specific queries into tool categories and extracting structured entities tailored to the requirements of each tool, all while operating under ambiguity and resource constraints. To enable retrieval, lookup, or reasoning operations that produce high-fidelity, actionable results, it is essential to accurately identify the right tool and precisely extract its required entities.

In the automotive domain, user queries span a wide array of scenarios. A user might request a technical service bulletin (TSB)\footnote{Technical Service Bulletins (TSBs) are documents issued by manufacturers that recommend procedures for repairing recurring, unanticipated vehicle issues.} for a known issue, seek safety recall\footnote{A car safety recall is issued when a manufacturer or regulatory agency (such as NHTSA in the U.S.) identifies a defect in a vehicle component that could pose a safety risk.} or customer complaint data, look up repair procedures from service manuals\footnote{Service manuals are official documents issued by vehicle manufacturers that contain detailed procedures and specifications for diagnosing, repairing, and maintaining vehicles.}, inquire about symptom-based diagnostics, search for parts needed for a specific repair, retrieve component-specific information such as part numbers or prices, or determine parts required for routine maintenance based on time or mileage. Each of these information needs corresponds to a distinct internal tool, optimized for that specific task. Descriptions and representative examples of these tool categories are presented in Table~\ref{tool-category-descriptions-examples}. Consequently, an effective query understanding system must satisfy three essential criteria: (1) accurately determine the most appropriate tool for a given query, (2) extract tool-specific structured entities necessary for downstream processing, and (3) operate with sufficiently low latency to support real-time deployment in production environments.

To illustrate the significance of precise tool selection and entity schema alignment, consider the following example query: "Replace brake pads for my Toyota Corolla 2015."  \\
This query should be directed to the \texttt{repair\_to\_parts} tool, which returns the parts required to perform the repair. The tool also necessitates structured input in the following format: \\
\{                                                   \\
\hspace*{2em}"tool\_category": "repair\_to\_parts",  \\
\hspace*{2em}"entities": \{                          \\
\hspace*{4em}"make": "Toyota",                       \\
\hspace*{4em}"model": "Corolla",                     \\
\hspace*{4em}"year": 2015,                           \\
\hspace*{4em}"component": "brake pads",              \\
\hspace*{4em}"labor\_action": "replace"              \\
\hspace*{4em}\}                                      \\
\}                                                   \\
Now consider a superficially similar query: "How to replace brake pads for my Toyota Corolla 2015." \\
Despite containing identical phrase "replace brake pads", the user's intent here is to get a procedural guide from a vehicle's service manual. Thus, here the suitable tool is \texttt{techdoc}, and the corresponding schema differs as shown below: \\
\{                                                 \\
\hspace*{2em}"tool\_category": "techdoc",          \\
\hspace*{2em}"entities": \{                        \\
\hspace*{4em}"make": "Toyota",                     \\
\hspace*{4em}"model": "Corolla",                   \\ 
\hspace*{4em}"year": 2015,                         \\
\hspace*{4em}"query\_type": "procedure",           \\
\hspace*{4em}"component": "brake pads",            \\
\hspace*{4em}"system": null                        \\
\hspace*{4em}\}                                    \\
\} 

\begin{table*}[h]
  \centering
  \begin{tabular}{p{0.15\textwidth} p{0.45\textwidth} p{0.35\textwidth}}
    \hline
    \textbf{Tool Category} & \textbf{Tool Description}   &   \textbf{Example Query}                                        \\\hline
    tsb                  &  Fetches official technical service bulletins (TSB) issued by OEMs for known issues associated with a specific vehicle.    &  Is there any TSB about spark plug fouling in the 2015 Subaru Forester?                      \\\hline
    nhtsa                &  Returns government-reported safety recalls and customer complaints related to a specific issue for a vehicle.             &  Are there any complaints or recalls about spark plug misfires in 2017 Kia Optima?           \\\hline
    techdoc              &  Provides OEM repair procedures and technical specifications (e.g., torque, capacity) for specific components or systems, based on user queries about "how to" perform a task or "what is" the specification from the service manual. & What is the torque spec for installing spark plugs on a 2016 Chevy Malibu?                                                  \\\hline
    smart\_insights      &  Offers diagnostic insights, causes, and possible repairs based on symptoms described by the user. &  My car has rough idle and stalling. What could be the issue in a 2020 Corolla?                                             \\\hline
    parts\_catalog       &  Retrieves parts information such as part numbers, prices, images, and PNC for a specified vehicle component.  &  Show me the part number and price for spark plugs for a 2019 Ford Fusion.                                        \\\hline
    repair\_to\_parts     &  Determines the parts needed for performing a specific repair on a vehicle. & What parts are needed to replace spark plugs in a 2018 Honda Accord?                                                                          \\\hline
    service\_to\_parts   & Identifies parts required for routine maintenance services based on time or mileage intervals.  & What parts do I need for the 30,000-mile service on my Toyota Camry?                                                          \\\hline
    others   & Handles queries that fall outside the scope of all defined tool capabilities, including vague, irrelevant, or unsupported requests.  & What are the negative aspects of choosing an aftermarket brake pad over an OEM part?              \\\hline
  \end{tabular}
  \caption{Tool categories with descriptions and example queries}
  \label{tool-category-descriptions-examples}
\end{table*}

These examples highlight several fundamental challenges in automotive query understanding:

\begin{itemize}
    \item \textbf{Lexical ambiguity}: Phrases such as “replace brake pads” may lead the model to misinterpret the user’s intent, despite subtle contextual cues.
    \item \textbf{Entity schema variability}: Each tool expects a unique schema of structured inputs; for instance, \texttt{repair\_to\_parts} requires a \texttt{labor\_action}, whereas \texttt{techdoc} necessitates a \texttt{query\_type} and \texttt{system}. As such, entity extraction must dynamically adapt to the schema associated with the predicted tool.
    \item \textbf{Lack of public datasets}: The niche nature of this domain precludes the availability of large, labeled datasets to train models for accurate classification and tool-specific extraction. Consequently, synthetic data was generated, followed by meticulous expert review cycles to ensure data integrity.
    \item \textbf{Error propagation}: Misclassifying a query or extracting incorrect entities may trigger the invocation of an inappropriate tool or yield misleading information, such as irrelevant TSBs or inaccurate repair procedures. This can adversely affect technicians' decisions, potentially resulting in improper servicing and risks to safety, compliance, and user trust.
    \item \textbf{Cascading failures}: An erroneous classification can cause subsequent entity extraction to be performed using the wrong schema, compounding the error across the pipeline.
    \item \textbf{Production readiness}: Systems intended for integration into real-world pipelines must exhibit both low latency and high accuracy across classification and extraction components.
\end{itemize}

Early experiments with single-step prompting—where classification and extraction were performed jointly using LLMs—revealed that combining multiple reasoning objectives within a single prompt leads to moderate accuracy and significant latency overhead. The model’s capacity is divided between disambiguating the intent and generating structured entities, often resulting in suboptimal outcomes for both. To address this, we propose a modular two-step architecture that separates classification and extraction into distinct, specialized subtasks. The first step employs a lightweight model for rapid and precise classification, while the second step leverages an LLM with a compact, schema-specific prompt for precise entity extraction. This decomposition not only improves inference speed but also enhances overall accuracy and robustness.

Our main contributions are threefold: (1) We release a large-scale, high-quality labeled dataset tailored to tool classification and schema-aware entity extraction in the automotive domain. (2) We propose and validate a modular, two-step architecture that significantly improves both accuracy and latency compared to monolithic single-step prompting. (3) We demonstrate the effectiveness and scalability of our approach through a low-latency implementation suitable for real-world automotive pipelines.

\section{Related Work}

Domain-specific query understanding involves adapting general language models to niche areas with specialized vocabularies. Continued pretraining on domain-specific data has shown significant performance gains \cite{dont-stop-pretraining}. Parameter-Efficient Fine-Tuning (PEFT) techniques, including adapter modules \cite{peft} and Low-Rank Adaptation (LoRA) \cite{lora}, significantly reduce computational demands while maintaining model performance, crucial for domain-specific applications.

Synthetic data generation addresses data scarcity, with methods like Easy Data Augmentation (EDA) enhancing robustness through simple text variations \cite{eda}. More sophisticated approaches, such as back-translation \cite{uda} and LLM-generated data \cite{sdg-llm}, further expand data augmentation capabilities.

Modular architectures using specialized, independently optimized modules often outperform monolithic approaches in structured extraction tasks \cite{apie}. Schema-aware extraction models like Unified Information Extraction (UIE)  \cite{uie} leverages schema-based prompts to guide structured information extraction dynamically, accommodating various extraction schemas within a single model. Similarly, GenIE \cite{genie} employs constrained decoding to ensure schema consistency during generative extraction.

Contrastive learning frameworks like SetFit \cite{setfit} effectively handle few-shot classification tasks without extensive prompt engineering, facilitating rapid industry deployment.


\section{Dataset}

\subsection{Dataset Curation}
We curated a high-quality dataset combining manually written and synthetic data to support both monolithic and modular model configurations. The dataset enables fine-tuning for the combined (classification + entity extraction) setting as well as for the decomposed two-step design, ensuring consistent evaluation across both paradigms. The test data was designed to evaluate generalization, robustness, and edge-case performance. Data generation was aided by few-shot prompting with LLaMA 3.3 70B Instruct\cite{llama3_3_70b}, and all synthetic outputs were manually reviewed to ensure correctness and coverage.

\begin{figure*}[h]
  \includegraphics[width=\textwidth]{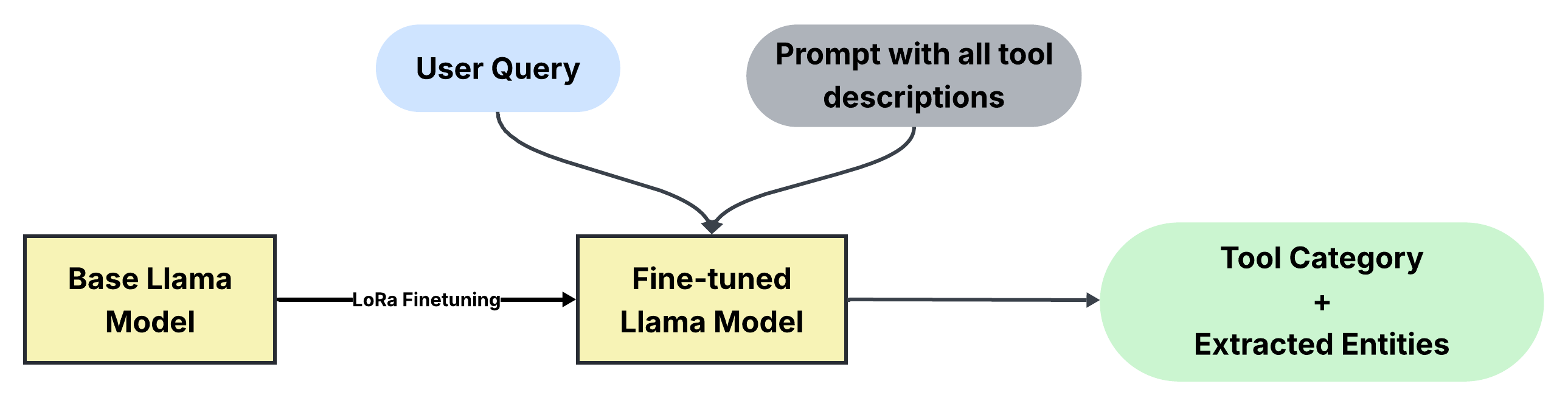}
  \caption{\textit{Single-step approach architecture.} A LLaMA 3.2 3B Instruct model is fine-tuned on a combined dataset (Section~\ref{sec:training-data}) designed for both tool classification and entity extraction. At inference time, a user query is directly passed to the fine-tuned model, which outputs a structured JSON response containing the predicted tool category and the corresponding extracted entities. }
  \label{single-step-approach-architecture}
\end{figure*}

\subsubsection{Training Data}
\label{sec:training-data}

\begin{itemize}
  \item \textbf{Dataset 01 (Manual Core Set):} A seed dataset of 216 samples was manually curated. Every sample was labeled with the correct tool and annotated with corresponding tool-specific entities. This dataset served as the basis for downstream data generation.
  \item \textbf{Dataset 02 (Manual + Reasoning):} To enhance interpretability, we expanded each sample in Dataset 01 by adding a short reasoning statement explaining why the query belongs to the assigned tool category. These rationales aim to capture the intent-tool mapping logic in natural language.
  \item \textbf{Dataset 03 (Classification + Extraction Synthetic Data): }The 216 samples from Dataset 01 were used in one-shot prompts to Llama 3.3 to generate 6,069 synthetic samples containing both tool and entity information. After being manually reviewed, the produced data was added to Dataset 01, yielding 6285 training samples in total. 
  \item \textbf{Dataset 04 (Manual Classification-Only Data for SetFit Training): }This manually curated dataset has five distinct versions (V1–V5) of small datasets tailored for SetFit-based\cite{setfit} classification training in the first stage of our two-step architecture. Each version of this dataset contains around 17-20 representative quality samples per tool category on average, with minor modifications across versions to refine coverage and improve representativeness. 
\end{itemize}

\subsubsection{Evaluation Data}

\begin{itemize}
  \item \textbf{Dataset 01 (Manual):} A reliable and representative test set was manually curated with a collection of 201 queries covering all the tools, and a variety of query formulations was produced.
  \item \textbf{Dataset 02 (Generated):} Using the 201 manually curated samples as few-shot examples, 280 synthetic test queries were generated with Llama 3.3 to evaluate performance on model-driven data while keeping the style and intent consistent with real user inputs.
  \item \textbf{Dataset 03 (Generated - Tricky Cases):} To test the model for the edge cases, 35 challenging sample queries were manually crafted and utilized as examples in a few-shot prompt to generate 130 test samples using Llama 3.3. This dataset targets scenarios where the correct tool is non-obvious or contextually nuanced.
  \item \textbf{Dataset 04 (Entity Extraction Evaluation Set):} A dedicated evaluation subset derived directly from the entity annotations in Training Dataset 01. This subset isolates the structured entity components required for each tool schema and was used exclusively to evaluate entity-extraction accuracy. No new data was generated; instead, this dataset repurposes the manually verified entity fields from the training core set.
\end{itemize}

\subsection{Human Annotation \& Data Review}
To ensure the accuracy and domain relevance of our training and evaluation datasets, we incorporated human supervision by automotive domain experts into our dataset curation and review process. A human expert wrote all of the content for Training Dataset 01, which was used as the basis for training and to direct the creation of synthetic data.  Additionally, reasoning was manually added to each instance for Dataset 02.

The synthetically generated datasets (Datasets 03 and 04) went through manual review by the domain expert to ensure factual correctness, right labels and consistency with domain-specific language and logic. For the evaluation of experiments, Dataset 01 was hand-picked to provide a high-quality standard benchmark. Evaluation datasets 02 and 03 were synthetically generated followed by manual review by the expert. This rigorous human-in-the-loop curation was instrumental in achieving high model accuracy and robustness, especially under domain-specific ambiguity and schema variability.

The human expert involved in these annotation and review processes possesses substantial domain knowledge, with two years of dedicated experience in automotive information systems and labeled data generation. Through continuous exposure to real-world query handling and structured data creation tasks in the automotive domain, the expert has developed a nuanced understanding of maintenance, repair, and diagnostics workflows, qualifying them to oversee data quality and domain alignment throughout the dataset lifecycle.

\subsection{Dataset Statistics}
We compiled and organized multiple datasets for training and evaluation, combining both manually curated and synthetically generated samples. Table~\ref{dataset-statistics} summarizes the key statistics, including the purpose, number of samples, and type. Synthetic datasets were generated using few-shot prompting and subsequently reviewed by a domain expert to ensure consistency and accuracy. This comprehensive dataset design enables controlled comparison between single-step and two-step approaches under identical data conditions.

\begin{table}[h]
  \centering
  \begin{tabular}{cccc}
    \hline
    \textbf{Purpose} & \textbf{Dataset} & \textbf{Type} & \textbf{Samples} \\
    \hline
    \multirow{4}{*}{Training} 
    &  Dataset 01 & Human & 216                                                                               \\
    &  Dataset 02 & Human & 216                                                                               \\
    &  Dataset 03 & Synthetic & 6285                                                                          \\
    &  Dataset 04 & Human & Avg. 133                                                                \\\hline
    \multirow{4}{*}{Evaluation}
    &  Dataset 01 & Human & 201                                                                               \\
    &  Dataset 02 & Synthetic & 280                                                                           \\
    &  Dataset 03 & Synthetic & 130                                                                   \\
    &  Dataset 04 & Human & 216                                                                      \\\hline
  \end{tabular}
  \caption{Dataset statistics including source, number of samples and usage.}
  \label{dataset-statistics}
\end{table}

\begin{figure*}[h]
  \includegraphics[width=\textwidth]{Two-Step-Approach.png}
  \caption{\textit{Two-step approach architecture.} First, a user query is passed to a SetFit-based classifier trained on manually curated classification data (Section~\ref{sec:training-data}) to predict the tool category. This predicted category is used to select a corresponding tool-specific prompt from a prompt pool. The prompt and user query are then passed to the LLaMA 3.2 3B Instruct model for entity extraction. The final output is a structured JSON containing the predicted tool category and extracted tool-specific entities. }
  \label{two-step-approach-architecture}
\end{figure*}

\section{Methodology}
This section outlines the experimental methodologies employed to enhance the interpretation of user queries within a production-grade automotive assistant system. The study began with an investigation of a monolithic single-step architecture utilizing the LLaMA 3.2 3B Instruct and LLaMA 3.1 8B Instruct models\cite{llama3_2_3b, llama3_1_8b}. This baseline approach treated the problem as a unified task, combining tool classification with entity extraction within a single few-shot prompt. While this configuration simplified the overall pipeline, it forced the model to handle multiple reasoning objectives simultaneously, leading to moderate accuracy and elevated latency. Efforts to optimize the single step formulation by reduce prompt length and subsequently decrease inference latency included fine-tuning the LLaMA 3.2 on over 6,000+ synthetically generated examples. Despite these efforts, latency improvements remained marginal and the model’s output consistency deteriorated, particularly under the multi-task burden.

To address these limitations, we developed a modular two-step architecture that explicitly separates classification and entity extraction into separate stages. This design enables the use of smaller, specialized models for each subtask and more compact prompts for LLM inference. In this architecture, tool classification is delegated to a fine-tuned stella\_en\_400M\_v5\cite{stellav5} model developed by NovaSearch, trained using the SetFit contrastive learning framework\cite{contrast_learn}. Once the appropriate tool is identified, entity extraction is performed by prompting LLaMA 3.2 with a concise, tool-specific few-shot prompt that aligns with the schema of predicted tool.

\begin{table*}[h]
  \centering
  \begin{tabular}{ll}
    \hline
    \textbf{Tool Category} & \textbf{Entities} \\
    \hline
    tsb                  &  make, model, year, issue                                                                    \\
    nhtsa                &  make, model, year, mileage, issue                                                           \\
    techdoc              &  make, model, year, query\_type, component, system                                           \\
    smart\_insights      &  make, model, year, mileage, issue                                                           \\
    parts\_catalog       &  make, model, year, component, brand, warranty, pnc                                          \\
    repair\_to\_parts     &  make, model, year, labor\_action, component                                                \\
    service\_to\_parts   & make, model, year, service\_name, service\_type, service\_unit, driving\_pattern             \\\hline
  \end{tabular}
  \caption{Tool categories and their corresponding entities}
  \label{tool-category-entities}
\end{table*}

\subsection{Motivation behind Modular Approach}
The motivation for transitioning from a monolithic single-step approach to a modular two-step design stems from both architectural limitations and deployment constraints. The single-step configuration detailed descriptions for all tools and their expected inputs into one prompt increased input complexity and response time, while also heightening the risk of hallucinations and schema misalignment.

Attempts to mitigate these issues through supervised fine-tuning\cite{sft} on synthetic datasets were partially effective. While fine-tuning improved generalization, the model’s performance suffered from structural inconsistencies, especially while jointly performing classification and extraction task. The joint formulation additionally limited flexibility as any modification to tool definitions required full retraining or revalidation of the end-to-end model.

In contrast, the two-step architecture offers modularity, interpretability and better resource efficiency. By isolating classification into a lightweight, prompt-free model (stella\_en\_400M\_v5) and reserving the large LLM exclusively for schema-aware entity extraction, the system achieves faster inference, easier debugging and higher stability. Each component can be optimized, evaluated, and iterated independently, making the modular approach a robust and maintainable choice for real-time automotive query understanding.

\subsection{Architecture}
We examine two architectural paradigms for automotive domain query understanding: (1) a single-step pipeline and (2) a modular two-step pipeline.

\subsubsection{Single-step Architecture}

In the single-step design, we first experimented with direct prompting of the LLaMA 3.1 8B and LLaMA 3.2 3B Instruct models to jointly perform tool classification and entity extraction. In this configuration, each model received a single composite prompt containing the user query along with descriptions of all available tools and their expected input schemas. The model was required to generate a structured JSON output specifying both the predicted tool and the corresponding entities. These initial experiments established baseline behavior and helped assess the model’s ability to handle joint reasoning without additional training.

Subsequently, the LLaMA 3.2 3B model was fine-tuned (Figure~\ref{single-step-approach-architecture}) using LoRA~\cite{lora} on a curated, domain-specific dataset to improve consistency and schema alignment. The training data jointly encoded tool classification and entity extraction objectives, allowing the model to learn the mapping between user intents and structured outputs. While this approach simplified the overall pipeline, it made performance highly sensitive to prompt structure and increased the model’s dependence on multi-task reasoning stability.

\begin{table}[h]
  \centering
  \begin{tabular}{p{5cm}p{2cm}}
    \hline
    \textbf{Hyperparameter} & \textbf{Value}  \\
    \hline
    LoRA rank ($r$) & 8 \\
    LoRA scaling factor ($\alpha$) & 16 \\  
    Batch size & 2 \\
    Gradient accumulation steps & 2 \\
    Number of epochs & 10 \\
    Learning rate & $2 \times 10^{-4}$ \\
    Weight decay & $1 \times 10^{-2}$ \\
    Max gradient norm & 0.3 \\
    Warm-up ratio & 0.1 \\
    Dropout & 0.05 \\
    Optimizer & PagedAdam (32-bit) \\
    \hline
  \end{tabular}
  \caption{\label{hyperparameters-single-step}
    Training hyperparameters for single-step approach
  }
\end{table}

\begin{table*}[h]
  \centering
  \begin{tabular}{ccccc}
    \hline
    \textbf{Model} & \textbf{Task} & \textbf{Evalaution Dataset} & \textbf{Accuracy} & \textbf{F1 Score} \\
    \hline
    \multirow{4}{*}{Llama 3.2 3B Instruct}
    & Classification & Dataset 01 & 88.00\% & 88.10\%   \\
    & Classification & Dataset 02 & 88.90\% & 89.80\%   \\
    & Classification & Dataset 03 & 86.90\% & 89.60\%   \\
    & Entity Extraction & Dataset 04 & 91.00\% & -   \\
    \hline
    \multirow{4}{*}{Llama 3.1 8B Instruct}
    & Classification & Dataset 01 & 91.50\% & 93.50\%   \\
    & Classification & Dataset 02 & 91.70\% & 93.70\%   \\
    & Classification & Dataset 03 & 91.50\% & 92.90\%   \\
    & Entity Extraction & Dataset 04 & 93.00\% & -   \\
    \hline
  \end{tabular}
  \caption{\label{baseline-results}
   Baseline single-step prompting results across all evaluation datasets.
  }
\end{table*}

\subsubsection{Two-step Architecture}

The two-step pipeline (Figure~\ref{two-step-approach-architecture}) decouples classification from entity extraction. First, a lightweight classifier based on stella\_en\_400M\_v5 is trained using the SetFit framework with contrastive learning on domain-specific examples. This classifier predicts the tool category for a user query without using a prompt. Based on this predicted tool, a corresponding tool-specific prompt is retrieved from a prompt pool. The user query and retrieved prompt are then passed to LLaMA 3.2 3B, which performs entity extraction. The final structured output combined the classification label and extracted entities. This architecture improved latency, enhanced modularity, and scalability for system evolution.

\subsection{Experiment Setup}
The task is partitioned into two subtasks: (1) identifying the correct tool category for a given query, and (2) extracting tool-specific entities based on the predicted category. Table~\ref{tool-category-entities} lists all tool categories and their associated entity schemas. All experiments were conducted on a single NVIDIA A100 GPU (40GB SXM4), 30 vCPUs, 200 GiB RAM, and 0.5 TiB SSD storage.

\subsubsection{Single-Step Approach}

The single-step pipeline uses the LLaMA 3.2 3B and LLaMA 3.1 8B in an end-to-end configuration, performing both tool classification and entity extraction within a single inference pass. LoRA fine-tuning was carried out on Llama 3.2 3B with 16-bit mixed precision, targeting attention layers (q\_proj, k\_proj, v\_proj, and o\_proj) and MLP layers (gate\_proj, up\_proj, and down\_proj). The optimal training hyperparameters identified through empirical tuning are listed in Table~\ref{hyperparameters-single-step}. Inference was conducted using a deterministic decoding setup ($temperature = 0.01$ and a $maximum\ token\ length = 1024$).

\subsubsection{Two-Step Approach}
This approach sequentially performs classification and entity extraction into two different steps.

\begin{itemize}
    \item \textbf{Step 1 - Classification: }
    A lightweight SetFit-based classifier was trained using contrastive learning. Several embedding models were evaluated as base encoder including sentence-transformers/paraphrase-mpnet-base-v2\cite{paraphrase-mpnet-base-v2}, sentence-transformers/all-roberta-large-v1\cite{all-roberta-large-v1}, sentence-transformers/all-MiniLM-L12-v2\cite{all-MiniLM-L12-v2} and NovaSearch/stella\_en\_400M\_v5 with the last achieving the most reliable performance. The key hyperparameters for training are summarized in Table~\ref{hyperparameters-two-step}.
    \item \textbf{Step 2 - Entity Extraction: }
    The LLaMA 3.2 model was used for tool-specific entity extraction, utilizing the same inference settings as in the single-step setup ($max\ tokens = 1024$, $temperature = 0.01$). A curated prompt pool supplied few-shot prompts tailored to the tool category predicted in Step 1.
\end{itemize}

\begin{table}[h]
  \centering
  \begin{tabular}{p{5cm}p{2cm}}
    \hline
    \textbf{Hyperparameter} & \textbf{Value}  \\
    \hline
    Training iterations\footnote{In SetFit, the number of training iterations refers to the number of contrastive examples generated per training instance.}  & 30 \\
    Learning rate & $2 \times 10^{-5}$ \\
    Warm-up ratio & 0.1 \\
    \hline
  \end{tabular}
  \caption{\label{hyperparameters-two-step}
    Training hyperparameters for classification step of two-step approach
  }
\end{table}

\begin{table*}[h]
  \centering
  \begin{tabular}{ccccc}
    \hline
    \textbf{Training Dataset} & \textbf{Evaluation Dataset} & \textbf{Latency (in seconds)} & \textbf{Accuracy} & \textbf{F1 Score} \\
    \hline
    Dataset 01 & Dataset 01 & 2.0873 & 66.10\% & 73.20\%   \\
    Dataset 02 & Dataset 01 & 2.1740 & 65.10\% & 71.00\%   \\
    Dataset 03 & Dataset 01 & 2.3824 & 63.10\% & 68.70\%   \\
    \hline
  \end{tabular}
  \caption{\label{single-step-approach-experiment}
    Performance of Single-Step Fine-Tuned LLaMA 3.2 3B Across Different Training Datasets 
  }
\end{table*}

\subsection{Evaluation Setup}
All evaluations were conducted using the same hardware configuration described in the experiment setup. All LLM evaluations were executed using the vLLM runtime, which offers optimized inference performance and efficient memory utilization.

To evaluate both the architectural variants, we employed Okareo, an AI evaluation platform designed for end-to-end testing and monitoring of language models in production-like environments. Okareo allows the definition of task-specific evaluation criteria, the uploading of test datasets as interactive scenarios, and detailed inspection of model behavior through intuitive dashboards. Okareo supports large-scale performance monitoring using features like integration-ready APIs, model-based validation, and scenario-based testing. These features enable us to evaluate the effectiveness of both open-ended and structured tasks, such as entity extraction and classification respectively, while keeping an eye on performance metrics and failure modes that are pertinent to production.

For the single-step approach, we used Okareo’s MULTI\_CLASS\_CLASSIFICATION test mode to measure the classification performance of the fine-tuned LLaMA 3.2 model. For the two-step architecture, the classification component—powered by the SetFit-trained Stella V5 model was evaluated using the same classification framework within Okareo where it provides standard classification metrics, including accuracy, precision, recall, and F1-score.

Entity extraction, being an open-ended generation task, required a more nuanced evaluation. Minor formatting variations in the structured JSON output were considered acceptable, whereas semantic errors were treated as critical failures. To support this, we utilized Okareo’s ModelBasedCheck module, which is a proprietary, general-purpose evaluation model capable of context-aware, reference-free scoring based on natural language prompts specifying the criteria for judgment. We configured this module with a custom prompt to assess the semantic correctness of extracted entities on a per-sample basis. Each output was marked as as binary pass or fail, which were then aggregated to compute the final accuracy metric for entity extraction.

\begin{table*}[h]
  \centering
  \begin{tabular}{ccccccc}
    \hline
    \textbf{Training Dataset} & \textbf{Evaluation Dataset} & \textbf{Batch Size} & \textbf{Epochs} & \textbf{Latency} & \textbf{Accuracy} & \textbf{F1 Score} \\
    \hline
    Dataset 04 - V1 & Dataset 01 & 16 & 1 & 0.0254 & 90.50\% & 91.60\%  \\
    Dataset 04 - V2 & Dataset 01 & 16 & 1 & 0.0252 & 97.00\% & 97.00\%  \\ 
    Dataset 04 - V3 & Dataset 01 & 16 & 1 & 0.0253 & 95.50\% & 95.50\%  \\ 
    Dataset 04 - V4 & Dataset 01 & 8  & 2 & 0.0260 & 97.00\% & 97.10\%  \\
    Dataset 04 - V5 & Dataset 01 & 8  & 2 & 0.0239 & 97.50\% & 97.50\%  \\
    \hline
  \end{tabular}
  \caption{\label{two-step-approach-experiment}
    SetFit Classification Accuracy and Latency (in seconds) Using Varying Training Datasets 
  }
\end{table*}

\begin{table*}[h]
  \centering
  \begin{tabular}{ccccccc}
    \hline
    \textbf{Training Dataset} & \textbf{Evaluation Dataset} & \textbf{Batch Size} & \textbf{Epochs} & \textbf{Latency} & \textbf{Accuracy} & \textbf{F1 Score} \\
    \hline
    Dataset 04 - V4 & Dataset 02 & 16 & 1 & 0.0235 & 99.60\% & 99.60\%  \\ 
    Dataset 04 - V4 & Dataset 03 & 16  & 1 & 0.0246 & 93.00\% & 93.10\%  \\
    \hline
  \end{tabular}
  \caption{\label{robustness-eval}
    Robustness evaluation of the SetFit classification model on two additional datasets. 
  }
\end{table*}

\section{Results \& Findings }
The primary objective of this work is to accurately classify user queries into predefined tool categories and extract tool-specific entities with high precision and low latency. We evaluate our proposed two-step architecture against two baselines - a single-step prompting method and a single-step fine-tuned model on the basis of classification accuracy, entity extraction accuracy and end-to-end latency, and overall robustness. Our findings illustrate the trade-offs between prompt-based reasoning, supervised fine-tuning, and modular decomposition, ultimately demonstrating the benefits of the proposed two-step approach for real-time, production-grade deployment.

\subsection{Baseline Single-Step Prompting}
We initially experimented with a single-step prompting approach using both the LLaMA 3.1 8B and LLaMA 3.2 3B models. In this configuration, a composite prompt contained descriptions of all tool categories along with few-shot examples for each. The models jointly performed tool classification and entity extraction within a single inference pass.

While this setup simplified the overall pipeline, it resulted in moderate accuracy and relatively high latency. Specifically, as summarized in Table~\ref{baseline-results}, the Llama 3.2 3B model achieved an average classification accuracy of 87.9\% and entity extraction accuracy of 91\%, whereas the larger Llama 3.1 8B achieved 91.5\% and 93\%, respectively. The larger model offered modest accuracy gains at the cost of substantially slower inference.

The latency comparison in Table \ref{latency-results} shows that even the smaller model requires nearly a second for an end-to-end generation, confirming that single-step prompting does not meet the responsiveness thresholds required for production-grade systems.

\begin{table*}[h]
  \centering
  \begin{tabular}{cp{1.8cm}p{2.8cm}l}
    \hline
    \textbf{Model} & \textbf{Architecture} & \textbf{Average Latency (in seconds)} & \textbf{Notes} \\
    \hline
    Llama 3.1 3B & Single-step & 0.845 &  Combined classification + extraction  \\
    Llama 3.1 8B & Single-step & 1.497 &  Combined classification + extraction  \\
    Stella V5 + Llama 3.2 3B & Two-step & 0.525 &  classification: 0.025 s, extraction: 0.5 s \\
    \hline
  \end{tabular}
  \caption{\label{latency-results}
   Average end-to-end inference latency (in seconds) per architecture. The corresponding server configurations and experimental setup for these measurements are described in Section~4.3.
  }
\end{table*}

\subsection{Single-Step Fine-Tuned Model}
To explore whether fine-tuning could mitigate the latency and consistency issues observed in the single-step prompting setup, we fine-tuned the LLaMA 3.2 3B Instruct model using LoRA-based supervised training. The motivation was to eliminate the need for few-shot examples and reduce the prompt length by omitting tool descriptions while preserving schema alignment. The training leveraged a combination of manually curated and synthetically datasets, as outlined in Section~\ref{sec:training-data}. The top-performing configurations are summarized in Table~\ref{single-step-approach-experiment}.

This approach successfully reduced prompt size but it introduced new challenges: (1) the model exhibited reduced consistency in generating structured JSON outputs, and (2) enlarging the synthetic dataset did not yield significant quality improvements, even when tool descriptions were reintroduced at test time. Increasing dataset size provided diminishing returns, suggesting that the coupling of classification and extraction tasks within a single reasoning step inherently limits performance and generalization.

\subsection{Two-Step Architecture}
To mitigate the drawbacks of the single-step designs, we developed a modular “divide-and-conquer” architecture that separates tool classification and entity extraction into independent components.

\subsubsection{Tool Classification with SetFit}
The classification component employs the Stella V5 (400M) model fine-tuned with the SetFit contrastive learning framework. SetFit derives high-quality sentence embeddings from limited data and trains a lightweight classification head on top. With 17–20 training examples per tool category, the classifier achieved an accuracy of 97.00\% and an average inference latency of 0.025 seconds (Table \ref{latency-results}). This latency remained consistent across both A100 and H100 GPU servers. These results feature the efficiency of the approach in low-resource settings and scalability for introduction of new tool categories with minimal additional annotation. Detailed classification results across  multiple training dataset variants are shown in Table~\ref{two-step-approach-experiment}.

The best-performing classifier was trained on Dataset 04-V4. Although other variants (V2, V4, V5) achieve comparable accuracy, Dataset 04–V4 yielded fewer misclassifications on straightforward queries, demonstrating superior robustness and generalization.

To further assess robustness, SetFit classifier was evaluated on two additional datasets - Dataset 02 and Dataset 03, detailed in Section~\ref{sec:training-data}. The difficulty level of Dataset 02 matches the training data but Dataset 03 includes challenging queries with potential category ambiguity. The classifier maintained strong performance across both, as shown in Table~\ref{robustness-eval}. These evaluations were conducted using the best-performing model trained on Dataset 04-V4, validating the stability of the SetFit approach in diverse settings.

\subsubsection{Entity Extraction with LLaMA 3.2 3B}
Following tool classification, entity extraction was performed using the LLaMA 3.2 3B Instruct model. In this stage, the prompt includes only entity definitions and few-shot examples specific to the predicted tool category, eliminating the need for irrelevant schema descriptions and reducing input length. This focused prompting yielded an accuracy of 98\% and average latency of 0.50 seconds on an A100 server and 0.28 seconds on an H100 server. These results indicate a 1.6×–2.8× speedup compared to single-step approaches while simultaneously improving extraction precision. The model consistently produced well-structured JSON outputs and demonstrated strong reliability.

\section*{Limitations}

While our two-step architecture achieves strong empirical performance, several limitations remain. The classification model relies on 17–20 labeled samples per tool category, which is sufficient for the current tool inventory but may require manual adjustment as the number of tools or query diversity expands. A second limitation arises from the pipeline’s cascading nature as errors made in the classification stage directly propagate to the entity extraction stage, leading to compounded failures.Without a fallback or confidence-based rejection mechanism, misclassifications in the first step can adversely affect the overall output quality.

The entity extraction model occasionally generates verbose responses or includes unnecessary contextual details, requiring post-processing to ensure schema conformity. Moreover, the SetFit-based classifier, while efficient and accurate, lacks a prompting mechanism or open-intent detection capability. As a result, it cannot reliably handle out-of-scope or ambiguous queries that fall outside the predefined tool categories.

Finally, although latency is substantially reduced compared to monolithic baselines, real-time scalability remains constrained by the need for LLM inference in the entity extraction step, especially under high-concurrency workloads.



\bibliography{custom}




\end{document}